\documentclass[10pt]{article}
\usepackage{epsf}

\author{
V. G. Baryshevsky\footnote{E-mail bar@inp.minsk.by},  ~
D. N. Matsukevich\footnote{E-mail mats@inp.minsk.by}}
\title{Time-reversal violating rotation of polarization plane of light in
gas placed in the electric field}

\begin{document}
\maketitle

\begin{abstract}
The rotation of polarization plane of light in a gas placed in the
electric field is considered. Different mechanisms that cause this
phenomenon are investigated. Angle of polarization plane rotation for
transition  $6S_{1/2} \rightarrow 7S_{1/2}$ in cesium
($\lambda=539$nm) is estimated. The possibility to observe this effect
in experiment is discussed.
\end{abstract}

\section{Introduction}
Violation of time reversal invariance was discovered more then 30
years ago in $K$ meson decay \cite{Cristenson,CPLEAR}. Up to now it
remains one of the great unsolved problems in elementary particle physics.
Many experiments were devoted to the search of any other manifestation
of time reversal noninvariance. Among them for example are measurements of
electrical dipole moment (EDM) of neutrons \cite{Neutron}, atoms and
molecules \cite{Tl-EDM,Cs-EDM,TlF-EDM}. No EDM
was found but these experiments impose strong restrictions
on theory. In particular search of EDM in heavy atoms set a
tight limits on parameters of electron - nucleon $PT$ violating interactions
and value of electron EDM.

At present more precise schemes of experiment are actively discussed.
One of them is the observation of the light polarization
plane rotation caused by pseudo-Zeeman splitting of magnetic sub-levels
of atom with nonzero EDM in electric field. This effect arises due to
interaction $W = - \vec d_a \vec E$ of atomic EDM with external electric
field \cite{Zeldovich,NMOE,Budker}.  We should note here that
discussions of these experiments \cite{Zeldovich,NMOE,Budker} take into account
only static EDM of atom. According to \cite{Bar-in-E,Bar_lanl} atom has another
$PT$ noninvariant characteristic that describes its
response to the external electric field.
It is a $P$ and $T$ - odd polarizability $\beta^{PT}_E$ that arises
in electric field due to interference of $PT$ - odd and Stark - induced
transition amplitudes. As was shown in \cite{Bar-in-E} $PT$ - odd
polarizability also leads to the rotation of photon polarization plane
and circular dichroism of atomic gas in external electric field. This
contribution should exist even in a hypothetical case when
atomic EDM in a ground and excited states are occasionally equal to zero
and pseudo-Zeeman splitting of atomic levels is absent. Unlike rotation
of polarization plane due to atomic EDM that manifest Macaluso-Corbino
dependence of angle on light frequency, rotation caused by  $\beta^{PT}_E$
is a kinematic analog of Faraday rotation in a Van-Vleck paramagnetic.

Moreover $PT$ noninvariant polarizability $\beta^{PT}_E$ cause
magnetization of atom in electric field \cite{Bar_lanl,Bar_magnet}.
This magnetization in turn induce
magnetic field $H_{ind}$. The energy of interaction of magnetic moment
of atom with this field is $W_H =  - \vec \mu_a \vec H_{ind}(E)$. Therefore
the total splitting of atomic levels that cause rotation of polarization
plane of light is equal to $W_H =  -\vec d_a \vec E - \vec \mu_a \vec H_{ind}(E)$.
This splitting appears even if atomic EDM $d_a$ is equal to zero.

In this paper the mechanisms of rotation of polarization plane of light
due to both $PT$ noninvariant electron-nucleon interactions and
nonzero electron EDM are considered. Estimates of expected
angle of polarization plane rotation near
highly forbidden magnetic dipole transition $6S_{1/2} \rightarrow
7S_{1/2}$ in cesium ($\lambda=539$ nm) are performed. Possible experimental
schemes to observe this rotation are discussed.

\section{PT - odd mixing}
We start with the simplest case. Let us place an atom in a ground state $s_{1/2}$ to the
electric field. If we take into account admixture of the nearest $p_{1/2}$ state
due to $P$ and $T$ odd interactions and interaction with the electric field, then
the wave function of atom takes the form.
\begin{equation}
|{ \widetilde s_{1/2} }> = \frac{1}{\sqrt{4 \pi}}
(R_0 (r) - R_1 (r) (\vec \sigma \vec n) \eta
- R_1(r) (\vec \sigma \vec n) (\vec \sigma \vec E) \delta   )
|\chi_{1/2} \rangle
\end{equation}
Here $ \vec \sigma$ - are the Pauli matrices, $\vec n = \vec r /r$  is the
unit vector along the direction of $\vec r$, $R_0$ and $R_1$
are radial parts of $s_{1/2}$ and $p_{1/2}$ wave functions respectively,
$ | \chi_{1/2} \rangle$ is the
spin part of wave function, $\eta$ and $\delta$ are the mixing coefficients
due to $P$ and $T$ noninvariant interactions and electric field respectively.

Let us consider orientation of electron spin in atom. In order
to find the spatial distribution of spin direction we can calculate
the matrix element of electron spin operator in respect
to the spin part of atomic wave function. Only
terms proportional to the product of electrical field strength $\vec E$ and
$PT$ odd mixing coefficient $\eta$ are important for our consideration because only
they cause $PT$ - odd rotation of polarization plane of light.
The change of spin direction due to
these terms is
\begin{eqnarray}
\Delta \vec s (\vec r) & = &
\frac{ \eta \delta }{8 \pi} R_1^2
\left\langle  \chi_{1/2} |
(\vec \sigma  \vec n) \vec \sigma (\vec \sigma  \vec n) (\vec \sigma  \vec E) +
(\vec \sigma  \vec E) (\vec \sigma  \vec n) \vec \sigma  (\vec \sigma  \vec n)
| \chi_{1/2} \right\rangle \nonumber \\
 & = &
\frac{\eta \delta R_1^2}{ 8 \pi}
\biggl( 4 \vec n (\vec n \vec E) - 2 \vec E \biggr)
\end{eqnarray}
The vector field $ 4 \vec n (\vec n \vec E) - 2 \vec E $ is shown in Fig. 1.
Since  $\Delta \vec s$ does not depend on initial direction of atomic spin,
this spin structure appears even in non-polarized atom. Let us note that
the spin vector averaged over spatial variables differs from zero and
is directed along the vector of electric field strength $\vec E$.
The photons with directions of angular moment parallel and antiparallel
to the electric field will interact with such spin structure in a different ways
causing rotation of polarization plane of light.

Polarization of atoms produce magnetic moment of gas \cite{Bar_magnet}.
Thus we have another interesting
$PT$ - odd effect. If we place gas in electric field, small magnetic field will appear.
Magnetic field in turn will interact with magnetic moment of atom giving another
contribution to the rotation of the polarization plane of light \cite{Bar_lanl}.

\begin{figure}%N F2
\epsfxsize =8.5cm
\centerline{\epsfbox{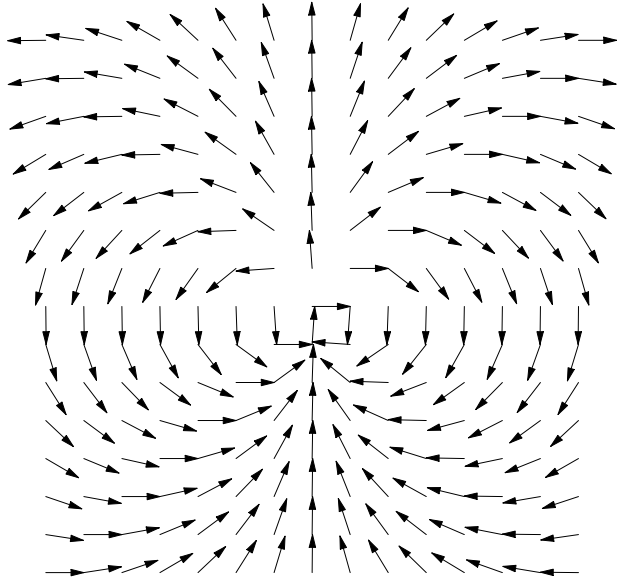}}
\vspace{7mm}
\caption{\it Vector field $4 \vec n (\vec n \vec E) - 2 \vec E$.
Vectors on figure shows direction of atomic spin in $s_{1/2}$ state if we take into
account admixture of $p_{1/2}$ state due to $PT$ noninvariant interactions
and external electric field.}
\vspace{0.5cm}
\label{figb}
\end{figure}

According to  \cite{Bar-in-E} if light propagates along the direction of
electric field, then the amplitude of elastic coherent forward
scattering of light by nonpolarized atoms has the form
\begin{equation}
f_{ik}(0) = f^{ev}_{ik} + \frac{\omega^2}{c^2}
(i \beta^P_s \epsilon_{ikl} n_{\gamma l}
+ i \beta^{PT}_{E} \epsilon_{ikl} n_{El})
\label{eq:amplitude}
\end{equation}
Here $f^{ev}_{ik}$ is  $P$ and $T$ invariant part of scattering amplitude,
$\beta^P_s$ is the $P$ - odd but $T$ -  even scalar atomic polarizability,
$\beta^{PT}_{E}$ is $P$ and $T$ -
odd scalar polarizability of atom, $\vec n_{\gamma} = \vec k / k$
is the unit vector along the direction of photon propagation,
$\vec n_{E} = \vec E / E$  is the unit vector along the direction of electric field,
$\epsilon_{ijk}$ is the third rank antisymmetric tensor.

The refraction index of gas can be written as
\begin{equation}
n=1+ \frac{2 \pi N}{k^2} f(0)
\label{eq:basicBar}
\end{equation}
where $N$ is the number of atoms per $cm^3$, $k$ is the photon wave vector.
Using (\ref{eq:amplitude}) expression (\ref{eq:basicBar}) can be rewritten as follows
\begin{equation}
n_{\pm} = 1+ \frac{2 \pi N}{k^2} (f^{ev} (0) \mp
\frac{\omega^2}{c^2} [\beta^P_s + \beta^{PT}_E (\vec n_E \vec n_{\gamma})] )
\label{eq:n_pm}
\end{equation}
Indices $+$ and $-$ stands for left and right circular polarization of incident
light respectively. Angle of rotation of polarization plane has the form
\begin{equation}
\phi = \frac{1}{2} k Re(n_{+} - n_{-}) = - \frac{2 \pi N \omega}{c}
(\beta^P_s + \beta^{PT}_E (\vec n_E \vec n_{\gamma}  ))
\label{eq:angle}
\end{equation}
Term proportional to $\beta^P_s$ describes well known phenomenon of $P$ -
odd but $T$ - even rotation of polarization plane of light. Term proportional
to  $\beta^{PT}_E$ describes  $P$ and $T$
noninvariant rotation of polarization plane of light about the direction of
electric field. $P$ and $T$ - odd rotation of polarization plane of light
change sign with the reversal of electric field direction in
contrast to $P$ odd but $T$ even rotation.
This allow us to distinguish $PT$ - odd rotation from the other
possible mechanisms of rotation of polarization plane.

Refraction index of gas has both real and imaginary parts. Since imaginary part of
refraction index for left and right circularly polarized photons are also different
due to $P$ and $T$ odd interactions, the admixture of circular polarization
to the linearly polarized light traveling in gas (circular dichroism) appears.

Let us consider $PT$ noninvariant polarizability $\beta^{PT}_E$.
According to \cite{Bar_lanl,Bar-in-E} the tensor of dynamical polarizability
of an atom (molecule) has the form
\begin{equation}
\alpha_{ik}^{n} = \sum_m
\Biggr\{
\frac{  \left\langle  {\widetilde g_n} |d_i| {\widetilde e_m} \right\rangle
      \left\langle  {\widetilde e_m} |d_k| {\widetilde g_n} \right\rangle
}{ E_{em}  - E_{gn} - \hbar \omega}
+
\frac{  \left\langle  {\widetilde g_n} |d_k| {\widetilde e_m} \right\rangle
      \left\langle  {\widetilde e_m} |d_i| {\widetilde g_n} \right\rangle
}{ E_{em}  - E_{gn} + \hbar \omega}
\Biggr\}
\label{eq:polarizability}
\end{equation}
where $|{\widetilde g_n}>$ and $|{\widetilde e_m}>$ are
wave functions of atom in ground and excited states perturbed by electric
field and $PT$ - noninvariant interactions,  $d$ is the operator of dipole transition,
$\omega$ is the frequency of incident light, $E_{em}$ and $E_{gn}$ are
the energies of atom in states $|{\widetilde g_n}>$ and $|{\widetilde e_m}>$
respectively.

In general case atoms are distributed to the sub - levels of ground state $g_n$
with the probability $P(n)$. Therefore, $\alpha _{ik}^{n}$ should
be averaged over all states $n$. As a result, the polarizability can be
written
\begin{equation}
\alpha_{ik} = \sum_{n} P(n) \alpha_{ik}^{n}  \label{10-1}
\end{equation}
The tensor $\alpha_{ik}$ can be decomposed into the irreducible parts as
\begin{equation}
\alpha_{ik} = \alpha_{0}\delta_{ik} + \alpha_{ik}^{s} + \alpha_{ik}^{a}.
\label{11}
\end{equation}
Here $\alpha _{0} = \frac{1}{3}{ \sum_{i} \alpha _{ii}}$ is the scalar,
$\alpha _{ik}^{s}=\frac{1}{2}(\alpha _{ik}+\alpha _{ki}) - \alpha_{0}\delta_{ik}$ is the
symmetric tensor of rank two, $\alpha_{ik}^{a}= \frac{1}{2}(\alpha_{ik}-\alpha_{ki})$ is the
antisymmetric tensor of rank two,
\begin{equation}
\alpha_{ik}^a = \frac{\omega}{\hbar}
\sum_{n} P(n)
\sum_m
\Biggr\{
\frac{  \left\langle  {\widetilde g_n} |d_i| {\widetilde e_m} \right\rangle
      \left\langle  {\widetilde e_m} |d_k| {\widetilde g_n} \right\rangle
-
     \left\langle  {\widetilde g_n} |d_k| {\widetilde e_m} \right\rangle
      \left\langle  {\widetilde e_m} |d_i| {\widetilde g_n} \right\rangle
}{ \omega^2_{em,gn}  -  \omega^2 }
\Biggr\}
\label{eq:antisym}
\end{equation}
where $\omega_{em,gn}=(E_{em}-E_{gn}) / \hbar $.

Let atoms (molecules) be nonpolarized. The antisymmetric part of
polarizability (\ref{eq:antisym}) is equal to zero in the absence of T- and P- odd
interactions \cite{Bar_lanl}. It should be reminded that
for the P-odd and T-even interactions the antisymmetric part of
polarizability differs from zero only for both the electric and magnetic
dipole transitions \cite{Bar_PT}.

We can evaluate the
antisymmetric part $\alpha _{ik}^{a}$ of the tensor $\alpha _{ik}$ of
dynamical polarizability of atom (molecule), and, as a result, obtain the
expression for $\beta _{E}^{PT}$\ in the following way. According to (\ref{eq:n_pm})
the magnitude of the $PT$-odd effect is determined by the polarizability
$\beta_{E}^{PT}$\ or by the
amplitude $f_{\pm }(0)$\ of elastic coherent scattering of a photon by an
atom (molecule). If $\overrightarrow{n}_{E}\parallel \overrightarrow{n}_{\gamma }$
the amplitude $f_{\pm }(0)$ in the dipole approximation can be
written as $f_{\pm }=\omega^2 \alpha_{ik} e^{(\pm)}_i e^{*(\pm)}_k /c^2
= \mp \omega ^{2} \beta _{E}^{PT} / c^2$. As a
result, in order to obtain the amplitude $f_{\pm }$, the polarizability
(\ref{eq:polarizability}) for photon polarization states
$\overrightarrow{e}= \overrightarrow{e}_{\pm }$ should be found.
 Using (\ref{eq:polarizability}) we can present the polarizability
$\beta_{E}^{PT}$ as follows:
\begin{equation}
\beta_{E}^{PT}=\frac{\omega }{\hbar }\sum_{n}P(n)
\sum_{m}
\Biggr\{
\frac{  \left\langle  {\widetilde g_n} |d_{-}| {\widetilde e_m}
\right\rangle \left\langle  {\widetilde e_m} |d_{+}| {\widetilde g_n}
			\right\rangle - \left\langle  {\widetilde g_n} |d_{+}|
		 {\widetilde e_m} \right\rangle \left\langle  {\widetilde e_m}
			|d_{-}| {\widetilde g_n} \right\rangle }{ \omega^2_{em,gn}  -
\omega^2 } \Biggr\}
\label{eq:beta_pt}
\end{equation}
For further
analysis the more detailed expressions for atom (molecule) wave
functions are necessary. The constants of $PT$ noninvariant
interactions are very small.  Therefore we can
use the perturbation theory. Let $|\overline g \rangle$ and
$|\overline e \rangle$ be the wave function of ground and excited
states of atom (molecule) interacting with an electric field
$\vec E$\ in the absence of $PT$ - odd interactions.  Switch on
$PT$ noninvariant interaction $(H_T\neq 0)$. According to the
perturbation theory the wave functions $|\widetilde g \rangle$
and $|\widetilde e \rangle$  can be written in this case as
\begin{eqnarray}
|{\widetilde g}> & = &
|\overline g> +
\sum_n |n> \frac{\langle n |H_T| \overline g \rangle}{E_g - E_n}
\nonumber \\
|{\widetilde e}> & = &
|\overline e> +
\sum_n |n> \frac{\langle n |H_T| \overline e \rangle}{E_e - E_n}
\label{eq:wave1}
\end{eqnarray}
where $H_T$ is Hamiltonian of $T$ noninvariant interactions.

It should be mentioned that both numerator and denominator of
(\ref{eq:beta_pt}) contain $H_T$. Suppose $H_T$ to be small one
can represent the total polarizability $\beta_{E}^{PT}$\ as the sum
of two terms

\begin{equation}
\beta _{E}^{T}=\beta_{mix}^{T}+\beta _{split}^{T}.  \label{sum}
\end{equation}
Here
\begin{equation}
\beta_{mix}^{PT}=\frac{\omega }{\hbar }\sum_{n}P(n)
\sum_{m}
\Biggr\{
\frac{
\left\langle  {\widetilde g_n} |d_{-}| {\widetilde e_m} \right\rangle
\left\langle  {\widetilde e_m} |d_{+}| {\widetilde g_n}	\right\rangle
-
\left\langle  {\widetilde g_n} |d_{+}| {\widetilde e_m} \right\rangle
\left\langle  {\widetilde e_m} |d_{-}| {\widetilde g_n} \right\rangle
}
{  \omega^2_{\overline{e}m,\overline{g}n}  - \omega^2 } \Biggr\}
\label{eq:beta_mix}
\end{equation}
where $ \omega_{\overline{e}m,\overline{g}n} $ does not include the $PT$ noninvariant shift of
atomic levels, and
\begin{equation}
\beta_{split}^{PT}=\frac{\omega }{\hbar }\sum_{n}P(n)
\sum_{m}
\Biggr\{
\frac{
\left\langle  {\overline g_n} |d_{-}| {\overline e_m} \right\rangle
\left\langle  {\overline e_m} |d_{+}| {\overline g_n} \right\rangle
-
\left\langle  {\overline g_n} |d_{+}| {\overline e_m} \right\rangle
\left\langle  {\overline e_m} |d_{-}| {\overline g_n} \right\rangle }
{ \omega^2_{em,gn}  - \omega^2 } \Biggr\}
\label{eq:beta_split}
\end{equation}
$$
\omega_{em,gn} = ( E_{em}(\vec E) - E_{gn} (\vec E))/ \hbar
$$
It should be reminded that
energy levels $E_{e,m}(\vec E)$ and $E_{g,n}(\vec E)$ contain
shifts caused by
interaction of electric dipole moment of atom with electric
field $\overrightarrow{E }$ and magnetic moment of atom with
T-odd induced magnetic field
$ \overrightarrow{H}_{ind}(\overrightarrow{E})$.

Below we will consider nonpolarized atoms and small detuning of
radiation frequency from atomic transition.
Therefore (\ref{eq:beta_mix}) and (\ref{eq:beta_split}) can be written as
follows.
\begin{equation}
\beta_{mix}^{PT}=\frac{1 }{2 \hbar (2j_g+1)}\sum_{n,m}
\Biggr\{
\frac{
\left\langle  {\widetilde g_n} |d_{-}| {\widetilde e_m} \right\rangle
\left\langle  {\widetilde e_m} |d_{+}| {\widetilde g_n}	\right\rangle
-
\left\langle  {\widetilde g_n} |d_{+}| {\widetilde e_m} \right\rangle
\left\langle  {\widetilde e_m} |d_{-}| {\widetilde g_n} \right\rangle
}
{  \omega_{\overline{e}m,\overline{g}n}  - \omega } \Biggr\}
\label{eq:beta_mix1}
\end{equation}
\begin{equation}
\beta_{split}^{PT}=\frac{1 }{2 \hbar (2 j_g + 1)}
\sum_{m,n}
\Biggr\{
\frac{
\left\langle  {\overline g_n} |d_{-}| {\overline e_m} \right\rangle
\left\langle  {\overline e_m} |d_{+}| {\overline g_n} \right\rangle
-
\left\langle  {\overline g_n} |d_{+}| {\overline e_m} \right\rangle
\left\langle  {\overline e_m} |d_{-}| {\overline g_n} \right\rangle }
{ \omega_{em,gn}  - \omega } \Biggr\}
\label{eq:beta_split1}
\end{equation}
In this section we will study only rotation of polarization
plane associated with $\beta_{mix}$. The rotation associated with
$\beta_{split}$ will be considered in the next section.

Due to Doppler shift resonance frequency of transition for a single
atom depends on velocity of atom in a gas.  In order to obtain
expressions for absorption length and angle of polarization plane
rotation we should average (\ref{eq:beta_mix}) over Maxwell
distribution of atomic velocity.  After standard calculations (see
e. q. \cite{Khriplovich}) expressions take the form.
\begin{eqnarray}
\phi  =
\frac{2 \pi N \omega}{c} \langle {\rm Re} \beta^{PT}_{mix} \rangle_{v} & = &
 - \pi N l \frac{\omega }{\Delta_{D} \hbar c} g(u,v)
[ |\overline{A^{+}}|^2 - |\overline{A^{-}}|^2]
\nonumber \\
L^{-1} =
2k \langle {\rm Im} n_{\pm} \rangle_v &=&
4 \pi N \frac{\omega }{\Delta_{D} \hbar c} f(u,v) |\overline{A^{\pm}}|^2
\label{eq:L_and_fi}
\end{eqnarray}
where $\langle  \rangle_v$ denotes the averaging
over atomic velocity,
$|\overline{A^{+}}|^2$ and $|\overline{A^{-}}|^2$ are the squares
of transition amplitudes for left and right
circularly polarized photons averaged over atomic polarization.
\begin{equation}
 |\overline{A^{\pm}}|^2 = \frac {1}{(2 j_g+1)}
\sum_{m_g}
\langle {\widetilde g} | d^{\pm} | {\widetilde e} \rangle
\langle {\widetilde e} | d^{\mp} | {\widetilde g} \rangle
\label{eq:A_pm}
\end{equation}
$\Delta_{D}=\omega_0 \sqrt{2 k T / M c^2} $ is Doppler linewidth,
$f(u,v)$ and $g(u,v)$ are equal to
\begin{equation}
\begin{array}{l}
g(u,v) \\
f(u,v)
\end{array}
\Biggr\}
=
\begin{array}{l}
{\rm Im} \\
{\rm Re}
\end{array}
\Biggr\}
\sqrt{\pi }e^{-w^{2}}\left( 1-\Phi (-iw) \right)
\end{equation}
here $w=u+i v$, $\Phi(z) = \frac{2}{\sqrt{\pi}} \int^{z}_{0} dt e^{-t^2}$,
$u=(\omega -\omega_0 )/\Delta_D $, $v=\Gamma / 2\Delta_{D} $, $\Gamma $ is
the recoil linewidth,

Let us assume that electric field is small enough and we can use first
order perturbation theory. Perturbed states
$|{\widetilde g}>$ and $|{\widetilde e}>$ in this case have the form
\begin{eqnarray}
|{\widetilde g}> & = &
|g> + \sum_n |n> \frac{\langle n |H_T| g \rangle}{E_g - E_n} +
\sum_m |m> \frac{\langle m | - \vec d \vec E_z |g \rangle}{E_g - E_m}
\nonumber \\
|{\widetilde e}> & = &
|e> + \sum_n |n> \frac{\langle n |H_T| e \rangle}{E_e - E_n} +
\sum_n |m> \frac{\langle m |  - \vec d \vec E_z |e \rangle}{E_e - E_m}
\label{eq:eANDg_bar}
\end{eqnarray}
Here $H_T$ is Hamiltonian of $PT$ noninvariant interactions,
 $| g >$ and $| e>$ are unperturbed ground and excited states of atom.
Only terms proportional to products of $H_T$ and $-\vec d \vec E$ leads
to phenomenon of interest.

Using (\ref{eq:eANDg_bar}) we can write.
\begin{equation}
|\overline{A^{+}}|^2 -|\overline{A^{-}}|^2  =
\frac{2}{2 j_g +1} {\rm Re} \sum_{m_g} (
\left\langle g | d^{PT}_{+} | e \right\rangle
\left\langle e | d^{St}_{-} | g \right\rangle -
\left\langle g | d^{PT}_{-} | e \right\rangle
\left\langle e | d^{St}_{+} | g \right\rangle )
\label{eq:A_pm2}
\end{equation}
where $d^{PT}_i$ is the admixture of $E1$ amplitude due to $PT$ - odd interactions
\begin{equation}
\left\langle g | d^{PT}_{i} | e \right\rangle =
\sum_m \frac {\left\langle g | H_T | m \right\rangle
\left\langle m | d_{i} | e \right\rangle}
{E_m - E_g} +
\frac {\left\langle g | d_{i} | m \right\rangle
\left\langle m | H_T  | e \right\rangle}
{E_m - E_e}
\end{equation}
and $d^{St}_{i} = \Lambda_{ik} E_k$ is the Stark - induced amplitude.
\begin{equation}
\left\langle g | d^{St}_{i} | e \right\rangle = E_k \Lambda_{ik} =
E_k \sum_n \frac {\left\langle g | d_k | n \right\rangle
\left\langle n | d_{i} | e \right\rangle}
{E_n - E_g} +
\frac {\left\langle g | d_{i} | n \right\rangle
\left\langle n | d_{k}  | e \right\rangle}
{E_n - E_e}
\end{equation}
Here $\Lambda_{ik}$ is the tensor of transition atomic polarizability.

According to \cite{Kozlov} Stark - induced $E1$ amplitude has the form
\begin{equation}
\left\langle e | d^{St}_{\epsilon} | g \right\rangle =
\sum_{q,q'} \Lambda_{q,q'} E_{-q} \epsilon_{-q'} =
\sum_{K,Q} (-1)^Q \Lambda^K_Q (E \otimes \epsilon)^K_{-Q},
\label{eq:Dst_lambda}
\end{equation}
where $E_q$ is external electric field strength,
$\epsilon_{q'}$ is the strength of electric field in a laser wave,
$\Lambda^K_Q$ and $(E \otimes \epsilon)^K_{-Q}$ are the components of irreducible
spherical tensors.

Using Wigner - Ekhard theorem we can represent  $\Lambda^K_Q$ as follows.
$$
\Lambda^K_Q = (-1)^{j_e-m_e}
\left( \matrix{
j_e  &  K &  j_g \cr
-m_e &  Q &  m_g \cr} \right)
\Lambda^K
$$
Reduced matrix elements  $\Lambda^K$ ($K = 0, 1, 2$) are proportional
to scalar, vector and tensor transition polarizability
respectively. Due to orthogonality of $3j$-symbols only terms
proportional to the vector part remains in (\ref{eq:A_pm2})
after summation over magnetic sub-levels.

In order to obtain the angle of polarization plane rotation and
absorption length for atoms with the nuclear spin,
we should take into account hyperfine structure of atomic levels.
After necessary transformations (see e. q. \cite{Khriplovich})
equation (\ref{eq:L_and_fi}) can be rewritten using reduced matrix
elements of corresponding transitions.
\begin{eqnarray}
\phi  &=&
- 4 \pi N_F l \frac{\omega}{\hbar c \Delta_D} g(u,v)
\frac{1}{3 (2F_g + 1) } K^2
Re (\left\langle g || d^{PT}  || e \right\rangle
    \left\langle e || d^{St}  || g \right\rangle )
\nonumber \\
L^{-1} & =& 4 \pi N_{F} \frac{\omega}{\hbar c \Delta_D}  f(u,v)
\frac{1}{3 (2F_g + 1) } K^2
| \left\langle g || A || e \right\rangle |^2
\label{eq:angle1}
\end{eqnarray}
Here $F_g$, $F_e$ are total angular moments of atom in ground and exited states,
$j_g$, $j_e$ are the total angular moments of electrons in these states,
$i$ is the nuclear spin,
$$N_F =N \frac{2F_g+1}{(2i+1)(2j_g+1)}$$ is the density of atoms with total moment $F_g$,
$$
K^2 = (2 F_g+1) (2 F_e+1)
\left\{ \matrix{
i  &  j_g &  F_g \cr
1  &  F_e &  J_e \cr} \right\}
$$
and  $\left\langle g || d^{St}  || e \right\rangle$ is proportional
to the $\Lambda^1$.
We assume that electric field is parallel to the direction of light
propagation.

\section{Rotation of polarization plane due to atomic EDM}
Presence of EDM in ground or excited state of atom also cause rotation
of polarization plane of light.
We can derive the expression for the angle of polarization plane
rotation performing the calculations similar to
those described in section 2, but using $\beta_{split}$ instead of
$\beta_{mix}$. But in this case the calculations can be greatly
simplified if we note that the mechanism of $PT$ noninvariant
rotation caused by atomic EDM is analogous
to the Faraday rotation of polarization plane in a week magnetic field.
Indeed according to \cite{Khriplovich, Faraday} application of weak
magnetic field to the atomic gas affect the refractive index in two ways:
through the changes in the energies of the magnetic sub-levels
and through the mixing of hyperfine states.

If we consider only terms proportional to the  magnetic field
strength $H$ and neglect the higher order terms
then the levels shift becomes \cite{Faraday}
$$
\Delta E_i = - H \langle i | \mu_z | i \rangle
$$
The magnetic field $H$ mixes states of
the same $F_z$ but different $F$, so the state $| j \rangle$
becomes
$$
| \overline{j} \rangle = | j \rangle -
\sum_{k \not= j  } H_z
\frac{| k \rangle \langle k | \mu_z | j \rangle} {E_k -E_j}
$$
If atom has an EDM then applied electric field affect the refraction index in
the same way (see  (\ref{eq:beta_split})). It shifts the atomic levels
$$
\Delta E_i = - E \langle i | d_z | i \rangle
$$
and mixes the hyperfine states of atom with the same $F_z$ but different $F$
$$
| \overline j \rangle = | j \rangle -
\sum_{k \not= j  } E_z
\frac{| k \rangle \langle k | d_z | j \rangle} {E_k -E_j}
$$
Therefore after substitutions $E \rightarrow H$,
$\mu_g  \rightarrow d_g$,
$\mu_e  \rightarrow d_e$
where $d_e$, $d_g$ are EDM of atom in
ground and excited states, $\mu_i$ is the magnetic moment of state $i$, we can
use in calculations the expression of \cite{Khriplovich,Faraday} for
rotation of polarization plane of light in a weak magnetic field.

If we take into account only
dipole transitions (it is possible for example for $6s_{1/2}
\rightarrow 7s_{1/2}$ transition in cesium), then the angle of
polarization plane rotation has the form
\begin{equation}
\phi = \frac{2 \pi N l}{(2 i+1)(2 j_g +1)} \frac{\omega}{\Delta_D \hbar c}
\frac{E_z}{\hbar \Delta_D} |A|^2 \biggl(
\frac{\partial g(u,v)}{\partial u} \delta_1  +
2 g(u,v) \gamma_1 \biggr).
\label{eq:dipole}
\end{equation}
where $A$ is the reduced matrix element of transition amplitude.
The expressions for parameters $\gamma_1$ and $\delta_1$
are given below
\begin{eqnarray}
\gamma_1 &=& \frac{(2 F_g + 1)(2 F_e + 1)}{\sqrt{6}} (-1)^i
\left\{ \matrix{
i  &  j_g &  F_g \cr
1  &  F_e &  j_e \cr} \right\}
[
d_e (-1)^{j_e + F_g} \sqrt{\frac{(j_e + 1)(2 j_e +1)}{j_e}}
\nonumber \\
 & & ( \frac{\Delta_D}{\Delta_{hf}(F_e, F_e -1)} (2 F_e -1)
\left\{ \matrix{
i  &  j_g    &  F_g \cr
1  &  F_e -1 &  j_e \cr} \right\}  % first 6j symbol
\left\{ \matrix{
i  &  j_e    &  F_e \cr
1  &  F_e -1 &  j_e \cr} \right\}  % second 6j symbol
\left\{ \matrix{
F_g  &  1    &  F_e \cr
1  &  F_e -1 &    1 \cr} \right\}  % third 6j symbol
\nonumber \\
& &
+
 \frac{\Delta_D}{\Delta_{hf}(F_e, F_e +1)} (2 F_e +3)
\left\{ \matrix{
i  &  j_g    &  F_g \cr
1  &  F_e +1 &  j_e \cr} \right\}  % first 6j symbol
\left\{ \matrix{
i  &  j_e    &  F_e \cr
1  &  F_e +1 &  j_e \cr} \right\}  % second 6j symbol
\left\{ \matrix{
F_g  &  1    &  F_e \cr
1  &  F_e +1 &    1 \cr} \right\} )  % third 6j symbol
\nonumber \\
& &  -
\biggl(
j_e \leftrightarrow j_g,
F_e \leftrightarrow F_g,
d_e \leftrightarrow d_g,
\biggr) ]
\nonumber
\end{eqnarray}
and
\begin{eqnarray}
\delta_1 &=& \frac{(2 F_g + 1)(2 F_e + 1)}{\sqrt{6}} (-1)^i
\left\{ \matrix{
i  &  j_g &  F_g \cr
1  &  F_e &  j_e \cr} \right\}^2
[
d_e (-1)^{j_e + F_g} \sqrt{\frac{(j_e + 1)(2 j_e +1)}{j_e}}
\nonumber \\
 & &  (2 F_e +1)
\left\{ \matrix{
i  &  j_e    &  F_e \cr
1  &  F_e    &  j_e \cr} \right\}  % first 6j symbol
\left\{ \matrix{
F_g  & F_e &  1   \cr
1    & 1   &  F_e \cr} \right\}  % second 6j symbol
+
\biggl(
j_e \leftrightarrow j_g,
F_e \leftrightarrow F_g,
d_e \leftrightarrow d_g,
\biggr)
]
\nonumber
\end{eqnarray}
Here $\Delta_{hf}$ is the hyperfine level splitting, $\Delta_D$ is the Doppler
linewidth, $j_g$ and $j_e$ are the
angular moments of electrons in atom, $i$ is the nuclear spin, $F_g$ and $F_e$
are total angular moments of atom in a ground and excited states respectively.

First term in  (\ref{eq:dipole}) arise from level splitting in electric field.
It describes effect similar to  Macaluso - Corbino rotation
of polarization plane in magnetic field.
Second term appears due to mixing of hyperfine levels with the different total moment
$F$ but the same $F_z$ in electric field.
It describes the $T$ noninvariant analog of polarization plane rotation due
to Van-Vleck mechanism.

\section{Estimates}
Let us compare the magnitude of $PT$ - odd polarization plane rotation for
different transitions. If spin of atomic nucleus is zero than the angle
of rotation of polarization plane per absorption length due to $PT$ - odd
level mixing according to (\ref{eq:L_and_fi}) has the form
\begin{equation}
\phi(L_{abs}) = \frac{g(u,v)}{4 f(u,v)}
\frac{ |\overline{A^{+}}|^2 - |\overline{A^{-}}|^2} {|\overline{A^{\pm}}|^2}
\end{equation}
If detuning $\Delta \sim \Delta_D$ then
$g \sim f \sim 1$.

As in the case of $P$ odd but $T$ even interaction
\cite{Khriplovich}, we will discuss the
rotation of polarization plane of light near magnetic dipole transitions.
We can estimate the angle of polarization plane rotation per absorption length
as follows.
\begin{equation}
 \phi( L_{abs} ) \sim \frac{(\left\langle
d \right\rangle^2 \left\langle  H_T \right\rangle / \Delta E)
(\left\langle  d \right\rangle E_z / \Delta E )}
{\left\langle  \mu \right\rangle^2 }
\sim \frac {\left\langle  H_T \right\rangle}
           {\alpha^2 \Delta E}
\frac{\left\langle d \right\rangle E_z  }{\Delta E}
\label{eq:M1_allowed}
\end{equation}
Here $\left\langle  d \right\rangle \sim e a_0$, $\left\langle  \mu
\right\rangle \sim \alpha \left\langle  d \right\rangle $ are
the values of $E1$ and $M1$ transition amplitudes,
$\left\langle H_T \right\rangle$ is the matrix element of $PT$ noninvariant interaction,
$\Delta E \sim Ry$ is the typical space between energies of opposite parity states,
$a_0$ is the Bohr radius,
$\alpha=1/137$ is the fine structure constant.

Near strongly forbidden magnetic dipole transitions
(e. q. $6s_{1/2} \rightarrow 7s_{1/2}$ in $Cs$)
Stark-induced $E1$ amplitude is several orders of magnitude greater than
$M1$ one. Absorption of light here depends primarily on
Stark-induced amplitude. We can write for angle of
polarization plane rotation per absorption length the following expression
\begin{equation}
 \phi( L_{abs} ) \sim
\frac{(\left\langle  d \right\rangle^2 \left\langle  H_T \right\rangle
/ \Delta E)
(\left\langle  d \right\rangle E_z / \Delta E) }
{ (\left\langle  d \right\rangle^2 E_z / \Delta E)^2 }
\sim \frac {\left\langle  H_T \right\rangle}
           { \Delta E}
\frac{\Delta E }{\left\langle d \right\rangle E_z }
\label{eq:M1_forbidden}
\end{equation}
Usually  ${\left\langle d \right\rangle E_z } / {\Delta E }$ is
less than $10^{-3} \sim 10^{-4}$.
Therefore angle of rotation per absorption length
is higher for strongly forbidden $M1$
transition.

Angle of rotation per absorption length near allowed $E1$ transition is
essentially lower than (\ref{eq:M1_allowed}) and (\ref{eq:M1_forbidden}).
\begin{equation}
\phi( L_{abs} )
\sim
\frac{(\left\langle  d \right\rangle^2 \left\langle  H_T \right\rangle
/ \Delta E)
(\left\langle  d \right\rangle E_z / \Delta E) }
{ \left\langle  d \right\rangle^2  }
\sim
\frac {\left\langle H_T \right\rangle}
           { \Delta E}
\frac{\left\langle d \right\rangle E_z  }{\Delta E}
\end{equation}
It should be noted here that the angle of rotation of polarization plane
per unit length has the same order of magnitude in all three cases.
The difference in angle of
rotation per absorption length is caused by different absorption of light
near the corresponding transition.

It is interesting to compare these estimates with the rotation of polarization
plane caused by nonzero EDM of atom. In the absence of hyperfine structure
the angle of rotation per
absorption length caused by splitting of magnetic sub-levels in electric
field can be estimated using (\ref{eq:L_and_fi}) and (\ref{eq:dipole}) as follows
\begin{equation}
\phi(L_{abs}) =
\frac{1}{2(2 j_g +1)}
\frac{E_z \delta}{f \hbar \Delta_D}
\frac{\partial g}{\partial u}
\sim
\frac{d_{at} E_z}{\hbar \Delta_D}
\sim
\frac{\left\langle d \right\rangle E}{\hbar \Delta_D}
\frac{\left\langle H_T \right\rangle}{\Delta E},
\end{equation}
where $d_{at} \sim \langle d \rangle \langle H_T \rangle / \Delta E$ is the EDM of atom,
$\Delta_D \sim (10^{-5} \sim 10^{-6}) \Delta E$ is the Doppler linewidth.
The value $\phi (L_{abs})$ here does not depend on magnitude of transition amplitude
$A$ and has the same order of magnitude for all kinds of transition considered above.

\section{P and T - odd interactions in atom}
Several mechanisms can cause the $PT$ noninvariant interactions in atom.
According to \cite{Khriplovich} they include $PT$ - odd weak interactions
of electron and nucleon, the interaction of electric dipole moment of
electron with the electric field inside the atom, interaction of
electrons with electric dipole and magnetic quadrupole moments of nucleus
and $PT$ odd electron - electron interaction.

Below we will consider two kinds of $PT$ - odd interactions that according
to \cite{Khriplovich} give the dominant contribution in our case.
This is $PT$ - odd electron nucleon interaction and interaction of electron
EDM with the electric field inside atom.

According to \cite{Khriplovich,Barr,Hunter} Hamiltonian of $T$ - violating
interaction between electron and hadron has the form
\begin{equation}
H_T = C_s \frac{G}{\sqrt{2}}(\bar e i \gamma_5 e) (\bar n n)
+ C_t \frac{G}{\sqrt{2}} (\bar e i \gamma_5 \sigma_{\mu \nu} e)
(\bar n \sigma^{\mu \nu} n)
\label{eq:hep_hamiltonian}
\end{equation}
where  $G=1.055 \cdot 10^{-5} m_p^{-2} $ is Fermi constant, $e$ and $n$
are electron and hadron field operators respectively, $C_s$ and $C_t$
are dimensionless constants that characterize the strengths of
$T$ - violating interactions relative to usual $T$ - conserving weak interaction.
The first term in (\ref{eq:hep_hamiltonian}) describes scalar hadronic
current coupling to pseudoscalar electronic current, and the second one
describes tensor hadron current coupling to the pseudotensor electron current.

Matrix elements for this $T$ - odd Hamiltonians according to \cite{Khriplovich}
is equal to
\begin{equation}
 \left\langle s_{1/2} || H_{\rm T odd} || p_{1/2} \right\rangle  =
 \frac{G m_e^2 \alpha^2 Z^2 R}{2 \sqrt{2} \pi}
\frac{{\bf Ry}}{\sqrt{ \nu_s \nu_p}^3} 2 \gamma C_s A
\label{Todd}
\end{equation}
where $m_e$ is the electron mass, ${\rm Ry}=13.6 \; {\rm eV}$  is Rydberg energy constant,
$\nu_i$ is the effective principal quantum number of state $i$,
$A$ is the atomic number,
$R$ is the relativistic factor ($R=2.8$ for cesium),
$\gamma = \sqrt{ (j+1/2)^2 -Z^2 \alpha^2 }$
and $j$ is the total angular moment of atom.
We neglect here tensor part of interaction for simplicity.

Hamiltonian of interaction of electron EDM and electric field inside atom
that mix opposite parity atomic states has the form \cite{Khriplovich}
\begin{equation}
H_d = \sum_k (\gamma_{0k} - 1) \vec \Sigma_k \vec E_k
\label{eq:EDM1}
\end{equation}
where $E_k$ is the electric field strength acting upon electron $k$.
When summation in (\ref{eq:EDM1}) is performed over one
valence electron only and electric field strength near the nucleus
approximately  equals to
$\vec E=Z \alpha \vec r / r^3$,
matrix element of operator $H_d$ can be written as follows \cite{Khriplovich}
\begin{equation}
\left\langle j, l=j+1/2 || H_d ||  j, l'=j-1/2 \right\rangle =
- \frac{4 (Z \alpha)^3 }{\gamma (4 \gamma^2 -1) (\nu_l \nu_{l'})^{3/2} a_0^2}
\label{eq:EDM_mat_el}
\end{equation}
where $l$ and $l'$ are the orbital angular moments, $a_0$ is the Bohr radius.

\section{Estimates for $6s_{1/2} \rightarrow 7s_{1/2}$ transition in cesium}
Let us estimate the $PT$ - odd rotation of polarization plane for
highly forbidden $M1$ transition $6s_{1/2} \rightarrow 7s_{1/2}$  in cesium.
The schemes of atomic levels of cesium is shown in Fig. 2.
\begin{figure}%N F2
\epsfxsize =6.5cm
\centerline{\epsfbox{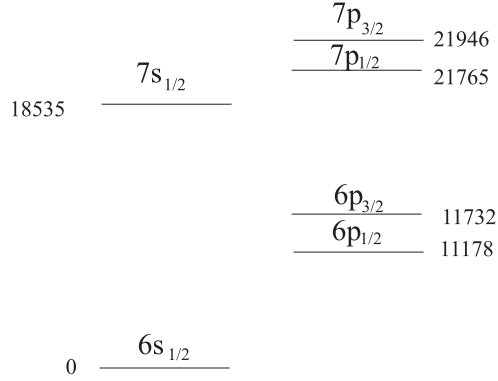}}
\vspace{7mm}
\caption{\it
Scheme of atomic levels for cesium. Energy of atomic levels
is given in $cm^{-1}$. }
\vspace{0.5cm}
\label{figc}
\end{figure}

\subsection{Rotation of polarization plane of light due electron nucleon interactions}

In order to obtain the angle of polarization plane rotation for the highly forbidden
$M1$ transition $6s_{1/2} \rightarrow 7s_{1/2}$ due to $P$ and $T$ noninvariant
interactions between electron and nucleus we can use the well known results
for $P$ - odd but $T$ - even weak interactions.
Matrix elements for $T$ - even Hamiltonian $H_{w}$ has the form
\cite{Khriplovich}.
\begin{eqnarray}
\left\langle s_{1/2} || H_w || p_{1/2} \right\rangle  & = &
i \frac{G m_e^2 \alpha^2 Z^2 R}{2 \sqrt{2} \pi}
\frac{{\bf Ry}}{\sqrt{ \nu_s \nu_p}^3} Q_w
\label{Teven}
\end{eqnarray}
$Q_w=-N + Z (1-4 \sin \theta_W)$ is the week nuclear charge,
$N$ and $Z$ are the number of neutrons and protons in nucleus respectively,
$\sin^2 \theta_W$ is the Weinberg angle.

Comparing this expressions with (\ref{Todd}) and using the wave functions of $6s_{1/2}$ and
$7s_{1/2}$ cesium states perturbed by $P$ - odd but $T$ - even interactions
 \cite{Khriplovich}
\begin{eqnarray}
|\widetilde{ 6s_{1/2}}> &=& |6s_{1/2}> +
i 10^{-11} \left( - \frac{Q_{w}}{N} \right)
(1.17 |6p_{1/2}> + 0.34 |7p_{1/2}>)
 \nonumber \\
|\widetilde{ 7s_{1/2}}> &=& |7s_{1/2}> +
i 10^{-11} \left( - \frac{Q_{w}}{N} \right)
(0.87 |6p_{1/2}> - 1.33 |7p_{1/2}>)
\end{eqnarray}
one can obtain atomic wave functions perturbed by $P$ and $T$ noninvariant
interactions
\begin{eqnarray}
|\widetilde{ 6s_{1/2}}> &=& |6s_{1/2}> +
10^{-11} \left( 2 \gamma \frac{A}{N} C_s \right)
(1.17 |6p_{1/2}> + 0.34 |7p_{1/2}>)
 \nonumber \\
|\widetilde{ 7s_{1/2}}> &=& |7s_{1/2}> +
10^{-11} \left( 2 \gamma \frac{A}{N} C_s \right)
(0.87 |6p_{1/2}> - 1.33 |7p_{1/2}>).
\label{Cs_state}
\end{eqnarray}
After simple calculations reduced matrix element of $PT$ odd $E1$
transition can be written as follows
\begin{eqnarray}
 \left\langle 6 s_{1/2} || d^{PT} ||7 s_{1/2} \right\rangle  &= &
- \sqrt{3} 10^{-11} \left( 2 \gamma \frac{A}{N} C_s \right)
\biggl(  0.87 \left\langle 6 s_{1/2} || d ||6 p_{1/2} \right\rangle
\nonumber \\
 &  & - 1.17 \left\langle 6 p_{1/2} || d ||7 s_{1/2} \right\rangle
-0.34 \left\langle 7 p_{1/2} || d ||7 s_{1/2} \right\rangle
\biggr)
\nonumber
\end{eqnarray}
Using values of radial integrals  \cite{Khriplovich}
$\rho(6s_{1/2},6p_{1/2})= -5.535$, $\rho(7s_{1/2},6p_{1/2})= 5.45$,
$\rho(7s_{1/2},7p_{1/2})= -12.30$ we obtain
\begin{equation}
\left\langle 6 s_{1/2} || d^{PT} ||7 s_{1/2} \right\rangle  =
1.27 \cdot 10^{-10} |e| a_0 C_s
\end{equation}

The matrix element of Stark - induced $6s_{1/2} \rightarrow 7s_{1/2}$
transition in cesium has traditionally been written in form \cite{Bouchiat}
$$
\langle 6s_{1/2}, m' | d_i^{St} | 7s_{1/2}, m \rangle =
\alpha E_i \delta_{mm'}  +
i \beta \epsilon_{ijk} E_j \langle m' | \sigma_k | m \rangle
$$
where $m$ and $m'$ are magnetic quantum numbers of ground and excited
states of cesium, $E_i$ is the electric field strength, $\sigma_k$ is
the Pauli matrix, $\alpha$ and $\beta$ are the scalar and vector transition
polarizability (aee also (\ref{eq:Dst_lambda})). The value of
$\left\langle 6 s_{1/2} || d^{St} ||7 s_{1/2} \right\rangle$
introduced in (\ref{eq:angle1}) can be expressed for cesium via the
vector transition polarizability $\beta$ as follows.
$\left\langle 6 s_{1/2} || d^{St} ||7 s_{1/2} \right\rangle  =
\sqrt{6} \beta E$. Value of $\beta$  is well known from theoretical
calculations \cite{beta} as well as from experiment \cite{beta2}.
According to \cite{beta} it is equal to $\beta = 27.0 a^3_0$.
Therefore
\begin{equation}
\left\langle 7 s_{1/2} || d^{St} ||6 s_{1/2} \right\rangle  =
1.28 \cdot 10^{-8} |e| a_0 E(V/cm)
\end{equation}

When temperature is $T=750 K$ pressure of $Cs$ vapor is $p=10$ kPa
\cite{BigBlue}, concentration of atoms is $N=  10^{18} cm^{-3}$
and Doppler linewidth is $\Delta_D / \omega_0 =  10^{-6}$.
For transition between hyperfine levels $F_g = 4 \rightarrow F_e =
4$, where coefficient $K^2$ is maximal ($K^2=15/8$) when detuning $\Delta \sim
\Delta_D$, $v = \Gamma / 2 \Delta_D \simeq 0.1 $
and $f \approx 1$, $g \approx 0.7$, absorption length in
longitudinal electric field $E$ is equal to
$L_{abs}= 7 \cdot 10^{10}/ E^2 (V/cm)$ (length is measured in centimeters)
and angle of $PT$ noninvariant rotation of polarization plane is
$$\phi = 1.0 \cdot 10^{-13} C_s l E .$$
If $E=10^4 { \rm V/cm}$
then $L_{abs} = 7 { \rm m}$. The best signal to noise ratio is achieved when
$l = 2 L_{abs}$ \cite{Khriplovich}.
In this case $|\phi| = 1.3 \cdot 10^{-6} C_s$.
The lowest limit to the parameters of electron-nucleon
interaction $C_s < 4 \cdot 10^{-7}$ was set in \cite{Tl-EDM}.
Corresponding limit to the angle of rotation of polarization plane is
$|\phi| < 0.5 \cdot 10^{-12} \;\rm rad.$

\subsubsection{Cesium EDM}
Using wave functions (\ref{Cs_state}) we can obtain EDM
in  $6s_{1/2}$ and $7s_{1/2}$ states of $Cs$
$$ d_{6s_{1/2}} = - 1.35 \cdot 10^{-10} C_s |e| a_0$$
$$ d_{7s_{1/2}} = - 4.39 \cdot 10^{-10} C_s |e| a_0 .$$
Under the same condition of experiment as before expression (\ref{eq:dipole})
yields the angle of polarization plane rotation due to level splitting in electric field
$$ |\phi_1| = 1.4 \cdot 10^{-24} l C_s E^3_z ({\rm V/cm}) < 8 \cdot 10^{-16} $$
and the angle of rotation due to hyperfine levels mixing
$$|\phi_2| = 2.1 \cdot 10^{-24} l C_s E^3_z ({\rm V/cm}) < 1.2 \cdot 10^{-15}.$$
(We assume here that for detuning  $\Delta \sim \Delta_D$ functions
$g(u,v) \simeq 0.7$, $\partial g(u,v)/ \partial u \simeq 1.1$).
These angles are two orders of magnitude lower than one
arising from interference of Stark-induced and
$PT$ noninvariant transition amplitudes.

\subsection{Rotation of polarization plane of light due to electron EDM}
If the $PT$ noninvariant interaction in atom is caused by
interaction of electron EDM with the electric field of nucleus
then wave functions of $6s$ and $7s$ states of cesium take the form
\begin{equation}
|{\widetilde ns_{1/2}}>  =
|ns_{1/2}> + \sum_m |mp_{1/2}>
\frac{\left\langle mp_{1/2} |H_d| ns_{1/2} \right\rangle}{E_g - E_n}
\end{equation}
where matrix element of operator $H_d$ is given in (\ref{eq:EDM_mat_el}).

We should take into account admixture of states
$6p_{1/2}$ and $7p_{1/2}$ to $6s_{1/2}$ and  $7s_{1/2}$ states of cesium.
After necessary calculations perturbed wave functions of atom can be written as follows
\begin{eqnarray}
|{\widetilde 6s_{1/2}}> & = &
 |{6s_{1/2}}> -  (35 |6p_{1/2}> + 10.5 |7p_{1/2}>) d_e/ (e a_0) \nonumber \\
|{\widetilde 7s_{1/2}}> & = &
|{7s_{1/2}}> + (27.7 |6p_{1/2}> - 36.2 | 7p_{1/2}>) d_e/ (e a_0 )
\label{eq:EDM2}
\end{eqnarray}
Using Eq. (\ref{eq:EDM2}) we can obtain reduced matrix element of electric dipole
transitions between ${\widetilde 6s_{1/2}}$ and ${\widetilde 7s_{1/2}}$ states
\begin{eqnarray}
\left\langle { 6s_{1/2}} || d^{PT} || { 7s_{1/2}} \right\rangle & = &
 \sqrt{2/3} d_e (35 \rho(6p_{1/2}, 7s_{1/2})
+10.5 \rho (7p_{1/2}, 7s_{1/2}) \nonumber \\
 & & + 27.7 \rho (6p_{1/2}, 6s_{1/2}) ) =- 72  d_e
\label{eq:matel_electron}
\end{eqnarray}
and electric dipole moment of cesium in ground state $d_{6s_{1/2}}$ and excited state
$d_{7s_{1/2}}$
\begin{eqnarray}
d_{6s_{1/2}} &=& 131 d_e \nonumber \\
d_{7s_{1/2}} &=& 400 d_e ,
\label{eq:EDM_electron}
\end{eqnarray}
where $d_e$ is the electron EDM.

As we mention before two effects induce $T$ - noninvariant rotation
of polarization plane in electric field.
First of them is the interference of $PT$ - odd and Stark induced transition
amplitudes and second is the interaction of atomic EDM with
electric field.  After substitution of (\ref{eq:matel_electron}) to equation
(\ref{eq:angle1})
one can obtain the angle of rotation due to interference of amplitudes
$|\phi| < 0.6 \cdot 10^{-12}$ under the same experimental conditions
as before.

The rotation due to atomic EDM is a sum of two contributions.
Using first term in Eq. (\ref{eq:dipole}) and Eq. (\ref{eq:EDM_electron})
one can obtain for the rotation induced by splitting of
magnetic sub-levels in electric field the angle
$|\phi_1| < 1.3 \cdot 10^{-15} $. The
mixing of hyperfine components (second term in equation
(\ref{eq:dipole})) gives the contribution
$|\phi_2| < 2 \cdot 10^{-15} $.

For estimates we use experimental limit on electron EDM from \cite{Tl-EDM}
$|d_e| < 4 \cdot 10^{-27} |e| \; {\rm cm}$.
Limits on angles quoted above are close to those obtained for $PT$ - odd
electron nucleon interactions.

\section{Discussion of experiment}

The simplest experimental scheme to observe the pseudo - Faraday
rotation of polarization plane of light in electric field consist of
cell with atomic gas placed in the electric field and sensitive polarimeter.
In the case of large absorption length one can place this cell in resonator or
delay line optical cavity to reduce the size of experimental setup
(see e. q. \cite{Axions}).

Several schemes was proposed to increase the sensitivity of
measurements.  One of them is based on the nonlinear magneto - optic
effect (NMOE) \cite{NMOE,Budker}.  Due to ultra - narrow width of
dispersion like shaped Faraday rotation caused by NMOE the
sensitivity of this experiments to the PT noninvariant rotation of
polarization plane (the change of rotation angle with the change of
applied electric field) is several orders of magnitude higher than in
the conventional scheme.  The authors of \cite{Budker} hopes to
achieve the sensitivity to the cesium EDM $d_{Cs} <
10^{-26} |e| $ cm.  The corresponding limit to the electron EDM
is $d_{Cs} < 10^{-28} |e| $ cm.

Even higher sensitivity can probably provide the method of measurements
of polarization plane rotation proposed in \cite{Bar_lanl}. This
method is based on observation of evolution of polarization of light
in a a cell with atomic vapor and amplifying media with inverse
population of atomic levels placed in the resonator. According to
\cite{Bar_lanl} the compensation of absorption of light in the cell
allows to increase the observed angle of polarization plane
rotation.

\section{Conclusion}
In the present article we have considered phenomenon of rotation of
polarization plane of light in gas placed in electric field. Calculations of angle
of polarization plane rotation were performed for  $6S_{1/2} \rightarrow 7S_{1/2}$
transition in atomic cesium.
Two mechanisms that cause this effect are considered. They are
interference of Stark-induced and $PT$ noninvariant transition amplitude
and atomic EDM. Both of them can be induced by $PT$ noninvariant interaction between
electrons and atomic nucleus and by interaction of electron EDM with electric field
inside atom.

For the highly forbidden $M1$ transition $6S_{1/2} \rightarrow 7S_{1/2}$ in
cesium we can expect the angle of polarization plane rotation per
absorption length due to $PT$ -
odd atomic polarizability $\beta^{PT}_{mix}$  $\phi < 10^{-12}$.
This angle is three orders of magnitude greater
than one caused by atomic EDM for this transition.

Angle of polarization plane rotation can be significantly greater for other
atoms, for example rare-earth elements, where additional amplification
goes from close levels of opposite parity.  The interesting example is
the transition $6s^2 \; {}^{1}S_0 \rightarrow 6s5d \; {}^{3}D_1$,
($\lambda = 408 {\rm nm}$)  in $Yb$ \cite{Yb}, where the value of $PT$ odd
angle can be two orders of magnitude higher then for cesium due to
larger $PT$ noninvariant amplitude.

Therefore we can hope that experimental measurement of described phenomenon
can achieve sensitivity in measurements of parameters of $PT$ noninvariant
interactions between electron and nucleus and electron EDM, comparable with
the current atomic EDM experiments.

\end{document}